\documentstyle[psfig]{mn}
\begin{document}
%
\newcommand{\mysize}{\tiny}
\newcommand{\cs}{\scriptstyle \rm }
\newcommand{\ha}{H$\alpha~$}
\newcommand{\hb}{H$\beta~$}
\newcommand{\hg}{H$\gamma~$}
\newcommand{\hd}{H$\delta~$}
\newcommand{\he}{H$\epsilon~$}
\newcommand{\etal}{{\it et al. }}
%
\newcommand{\arc}{$^{\prime\prime}$}
\newcommand{\kms}{km~s$^{-1}~$}
\newcommand{\skms}{km~s$\scriptstyle ^{-1}$}
\newcommand{\mpm}{$\pm$}
\newcommand{\ang}{\AA~}
\newcommand{\sang}{$\scriptstyle \rm \AA~$}
\newcommand{\msol}{M$_{\odot}$~} 
\newcommand{\rsol}{R$_{\odot}$~} 
\newcommand{\lsol}{L$_{\odot}$~} 
\newcommand{\mdot}{\.{M}~}
\newcommand{\sm}{\mbox{$\mu$m}}
\newcommand{\pyr}{yr$^{-1}$}
\newcommand{\mum}{$\mu$m~}
\newcommand{\vinf}{\mbox{$v_{\infty}$}~}
\newcommand{\Teff}{\mbox{$T_{\rm eff}$}~}
%
\newcommand{\Ctwo}{\mbox{C$_{2}$}}
\newcommand{\Cthree}{\mbox{C$_{3}$}}
\newcommand{\CHp}{\mbox{CH$^{+}$}}
\newcommand{\Htwo}{\mbox{H$_{2}$}}
\newcommand{\philband}{$\rm A^{1}\Pi_{u}-X^{1}\Sigma^{+}_{g}~$}
\newcommand{\swanband}{$\rm d^{3}\Pi_{g}-a^{3}\Pi_{u}~$ }
\newcommand{\redband}{$\rm A^{2}\Pi-X^{2}\Sigma^{+}~$}
\newcommand{\Cthreeband}{$\rm A^{1}\Pi_{u}-X^{1}\Sigma_{g}^{+}~$}
\newcommand{\sysch}{$\rm A^{1}\Pi_{u}-X^{1}\Sigma_{g}^{+}~$}
\newcommand{\eight}{8.7~\mum}
\newcommand{\ten}{ 10~\mum}
\newcommand{\twone}{21~\mum}


\title{Time resolved  spectroscopy of the post-AGB star HD~56126}

\author{Ren\'e D. Oudmaijer
                           $^{1}$ and
        Eric J. Bakker
                           $^{2,3}$ \\
$^{1}$ Kapteyn Astronomical Institute, P.O.Box 800, NL-9700 AV Groningen,
The Netherlands\\
e-mail: r.d.oudmaijer@astro.rug.nl\\
$^{2}$ SRON Laboratory for Space Research Utrecht,
Sorbonnelaan 2, NL-3584 CA Utrecht, The Netherlands \\
$^{3}$ Astronomical Institute, University of Utrecht,
Princetonplein 5, NL-3584 CC Utrecht, The Netherlands\\
e-mail: ebakker@fys.ruu.nl \\
}

\date{received June 24 1994, accepted June 27 1994}

\maketitle

\author{Ren\'e D. Oudmaijer and Eric J. Bakker}

\begin{abstract}
We have investigated the report of Tamura and Takeuti that the \ha
line of the F-type post-AGB star HD~56126 is variable on time scales of
minutes.  To this end, HD~56126 was observed on two occasions with the
William Herschel Telescope.  Seventeen, respectively thirty spectra were
taken within time span of 1.5 hours in order to detect any short term
variations.  We find that the \ha line profile changed
strongly over the two month interval, but no evidence is found for
short term variability.
The variability Tamura and Takeuti claim to find is probably due to the
low signal-to-noise in their spectra.


\end{abstract}

\section{Introduction}

HD~56126 (F5I, 7$^{h}$13$^{m}$25.3$^{s}$ +10$^{o}$05$^{\prime}$09\arc)
belongs to the class of the so-called high latitude supergiants that are
thought to be in the post-Asymptotic Giant Branch (post-AGB) phase of
their evolution.  These are stars with supergiant spectral types, but
are located at unexpectedly high Galactic latitudes.  After it was found
that most of these stars show far-infrared excess emission due to
thermally re-radiating dust (Parthasarathy and Pottasch 1986; Trams~\etal 1991;
Oudmaijer~\etal 1992; Kwok 1993) the post-AGB nature became
a widely accepted interpretation for these stars.  The infrared excess
is explained as being a remnant of the high mass-loss the stars
underwent during the AGB phase.  The evolved status makes the objects
less luminous and older compared to massive supergiants,
placing them closer to the plane and
allowing them more time to get there, solving the high latitude problem.

During the AGB phase, carbon rich material from the helium burning shell was
dredged up to the photosphere of HD~56126.  This is suggested by the
detection of the 3.3~$\mu$m, infrared ``PAH'' feature by Kwok~\etal (1990),
and the detection of C$_{2}$ and CN absorption lines in the optical
spectrum by Bakker~\etal (1995).  A study by Parthasarathy~\etal (1992)
showed that HD~56126 is metal deficient by a factor of ten with respect
to solar metalicities, although C, N, O and S appeared solar.  This
photospheric abundance pattern resembles that of interstellar gas, where
the metals have been condensed onto grains (Lambert~\etal 1988;
Bond 1991; Van Winckel~\etal 1992).

One of the main problems in the study of late stages of stellar
evolution of low to intermediate-mass stars is whether there is
mass-loss during the post-AGB phase, and if there
is, the magnitude of the mass-loss rate.  After the core mass, the
post-AGB mass-loss rate is the most important parameter governing the
time scale of the evolution of these objects from the
AGB to the Planetary Nebula (PN) phase.
For example the transition of a star with a core mass 0.546~\msol, having
a Reimers mass-loss of  $10^{-8}$~\msol~\pyr~ takes 100,000 years
(Sch\"onberner 1981 and 1983). A period in which the AGB wind has
dispersed into the interstellar medium long before the central star has
reached a temperature high enough to ionize its circumstellar shell and
become observable as a PN.  However, as pointed out by Trams~\etal
(1989), increasing the mass-loss rate to $5\times10^{-7}$~\msol~\pyr~
accelerates this transition to only 5000 years,
which would make a PN readily observable.
It is difficult however to determine post-AGB mass-loss rates. All
``classic'' tracers of mass-loss rates such as circumstellar CO millimeter
emission,
far infrared excesses etc. trace  the former AGB wind rather than the
present-day mass-loss in post-AGB stars.

Most of the post-AGB stars exhibit mild \ha emission.  Some objects
show strong P-Cygni type emission, while most of the stars have shell
type profiles.  In the recent literature the \ha emission has been
interpreted as the star undergoing a post-AGB mass-loss (Trams~\etal
1989; Parthasarathy~\etal 1992; Slijkhuis 1992).  In fact, the \ha
emission is, next to the near-infrared excess observed in some of the
objects (Trams~\etal 1989), the only diagnostic suggested for post-AGB
mass-loss.  Although a P-Cygni type emission can be understood with a
simple expanding wind model, the double peaked shell type profiles
require somewhat more complicated geometries as for example rotating
disks (Waters~\etal 1993).  The mass-loss interpretation gets more into
difficulties when considering the fact that many of the stars show
variable \ha profiles.  Some of the stars have shell type \ha lines
where the ratio between the blue and the red peaks varies on time scales
of months (Arellano Ferro 1985; Waelkens~\etal 1991, Waters~\etal
1993).  Only in the case of HR~4049 the \ha variability could be
correlated with radial velocity variations and the orbital phase of the
star in a binary system (Waelkens~\etal 1991).  Although the number of
binary stars in the sample of post-AGB stars under consideration is
relatively high (Waelkens and Waters 1993), not all stars that exhibit
\ha variations are yet identified as members of a binary system.

Another interpretation that can give rise to variable \ha emission of
post-AGB stars stems from  stellar pulsation  theory.  The F-G type
post-AGB stars are located in a part of the HR diagram where the
photosphere is subject to instabilities (Sasselov 1993; Zalewski 1992).
A consequence of such instabilities has been pointed out by
L\`ebre and Gillet (1991a+b, 1992) in their work on RV-Tauri and
W-Virginis stars, objects that resemble post-AGB stars very closely.
These stars show variable \ha emission during their pulsation cycle.
L\`ebre and Gillet showed that inward and outward moving shocks within
the photosphere can produce line profiles remarkably similar to those
observed in the post-AGB stars discussed above.

In this work we aim at confirming the rapid changes of the \ha
profile of HD~56126 described by Tamura and Takeuti (1993, hereafter TT93).
These authors reported \ha line profile variations on time scales of
the order of minutes.
If the  variability on minute time scales is real,
it will prove very hard to attribute the observed \ha emission to
mass-loss. In principle it may  be possible to have short ``puffs'' of
mass  ejected from the photosphere and falling back again,
giving rise to blue and red emission peaks respectively. Such ``puffs''
are then indicative of a turbulent photosphere, so  the mass-loss
and stellar pulsation  should be closely intertwined. It would then be more
reasonable to explain the rapid \ha variability in terms of pulsation
solely than a combination of mass-loss and pulsation together.
Realizing the important implications caused by the short term
variability reported by TT93, we decided to obtain a  set of high
quality spectra with short exposure times of HD~56126.  For this purpose
the Utrecht Echelle Spectrograph
mounted on the 4.2m WHT telescope on La
Palma was employed.

\section{Observations and Data reduction}

\subsection{Observations}

The first set of observations was carried out on December $21^{\rm st}$ 1993
in service time with the Utrecht Echelle Spectrograph (Unger 1994)
mounted on the Nasmyth platform of the 4.2 m William Herschel Telescope,
La Palma, Spain.  The weather was fair, but the seeing was larger than
2\arc, which, compared with the slit width of 1\arc, resulted in a
somewhat lower signal-to-noise of the spectra than could have been
possible.  The detector was a $1124 \times 1124$ TEK CCD.  Wavelength
calibration was performed by observing a Thorium-Argon lamp.  The
central wavelength of the setting was 5587~\AA, resulting in a
wavelength coverage from 4700--7200~\AA.  In order to make a
compromise between signal-to-noise and time resolution, it was decided
to make exposures with increasing integration times.

A second set of data was obtained during a run on the WHT in the night
of February 26-27 1994.
The settings were the same as for the December run, except for the
central wavelength (7127~\AA), resulting in a wavelength
coverage from 5500~\AA, -- 1~$\mu$m.
This time the  strategy was aimed at searching for time variations
within minutes. Thirty exposures of 60 seconds and 30 seconds
were taken consecutively.
The sky was cloudy, resulting in many telluric features in
the spectra.
The spectral resolution
as measured from telluric absorption lines (full width at half maximum)
is 8.5~\kms.
The logs of the observations and the resulting signal-to-noise ratios
(measured in the \ha order) are provided in Tables~\ref{art8tab1} and
\ref{art8tab2}.

\begin{table}
\caption{Log of the observations December 21 1993}
\label{art8tab1}
\centerline{\begin{tabular}{lrcr}
\hline
 Time    & $t_{\rm exp}$ & Airmass   & $SNR$\\
 U.T.    & [sec.]    &           &    \\
\hline
         &           &           &     \\
23:24:39 &  60       &  1.465    & 50  \\
23:27:58 &  60       &  1.446    & 52  \\
23:30:56 &  90       &  1.430    & 62  \\
23:37:41 &  90       &  1.394    & 63  \\
23:41:05 &  90       &  1.378    & 63  \\
23:44:40 & 180       &  1.361    & 87  \\
23:49:35 & 180       &  1.339    & 89  \\
23:54:29 & 180       &  1.318    & 83  \\
23:59:23 & 180       &  1.299    & 82  \\
00:05:25 & 240       &  1.276    & 98  \\
00:11:19 & 240       &  1.256    & 90  \\
00:17:10 & 240       &  1.237    & 99  \\
00:23:09 & 240       &  1.219    & 95  \\
00:29:12 & 240       &  1.202    & 103 \\
00:35:08 & 240       &  1.186    & 100 \\
00:41:11 & 240       &  1.171    & 90  \\
00:47:11 & 240       &  1.158    & 94  \\
         &           &           &     \\
\hline
\end{tabular}}
\centerline{The $SNR$ has been calculated in the line free region
 between 6533-6540~\AA}
\end{table}

\begin{table}
\caption{Log of the observations February 27 1994}
\label{art8tab2}
\centerline{\begin{tabular}{lrcr}
\hline
 Time      &  $t_{\rm exp}$   &Airmass & $SNR$ \\
 U.T.      &  [sec.]      &        &     \\
\hline
           &              &        &   \\
 23:30:35  &          60  &  1.133 & 58\\
 23:34:37  &          60  &  1.141 & 55\\
 23:37:39  &          60  &  1.147 & 53\\
 23:40:39  &          60  &  1.153 & 37\\
 23:43:52  &          60  &  1.160 & 47\\
 00:22:18  &          60  &  1.26  & 72\\
 00:25:39  &          60  &  1.280 & 60\\
 00:28:51  &          60  &  1.292 & 68\\
 00:31:54  &          60  &  1.303 & 55\\
 00:35:23  &          60  &  1.317 & 60\\
 00:38:37  &          60  &  1.331 & 48\\
 00:41:48  &          60  &  1.344 & 55\\
 00:44:58  &          60  &  1.359 & 73\\
 00:47:53  &          60  &  1.372 & 52\\
 00:56:58  &          60  &  1.417 & 64\\
 01:00:10  &          60  &  1.435 & 57\\
 01:03:23  &          60  &  1.453 & 60\\
 01:06:38  &          60  &  1.471 & 58\\
 01:10:06  &          60  &  1.492 & 53\\
 01:13:26  &          60  &  1.513 & 59\\
 01:16:39  &          60  &  1.534 & 50\\
 01:19:57  &          60  &  1.557 & 60\\
 01:23:08  &          30  &  1.580 & 45\\
 01:25:32  &          30  &  1.597 & 29\\
 01:27:57  &          30  &  1.616 & 33\\
 01:30:39  &          60  &  1.637 & 54\\
 01:33:32  &          60  &  1.660 & 63\\
 01:36:26  &          60  &  1.685 & 56\\
 01:39:21  &          60  &  1.711 & 59\\
 01:42:18  &          60  &  1.738 & 53\\
           &              &        &   \\
\hline
\end{tabular}}
\end{table}

\subsection{Data reduction}

The data were reduced using the standard Echelle reduction software in
the IRAF package.  The data were bias subtracted and divided by a
normalized flatfield frame.  After the flatfielding, the individual
orders were optimally extracted from the image frame, where, using
simple photon statistics, pixels are  weighted.


The spectra contained 44 orders in the December run, and 48 in the
February run, that were continuum normalized for
further analysis.
The wavelength calibration was done by fitting a 3$^{rd}$ order Legendre
polynomial in the dispersion direction and a 4$^{th}$ order polynomial in
the cross-order direction. The dispersion  of the fit through the more than
500 identified Th-Ar lines was 3.8~m\AA~ and 7~m\AA~ for the
December and February spectra respectively.
During the process of
extraction the signal-to-noise ratio of the pixels
was calculated with simple photon statistics. Checks
with   the actual observed
signal-to-noise of the data proved to be  correct  within 15\%.
The $SNR$ spectra calculated in this way will be used further in this study.

\section{Is there variability?}

\subsection{Time scales of minutes}
\label{art8secmin}

\begin{figure*}
\centerline{\hbox{\psfig{figure=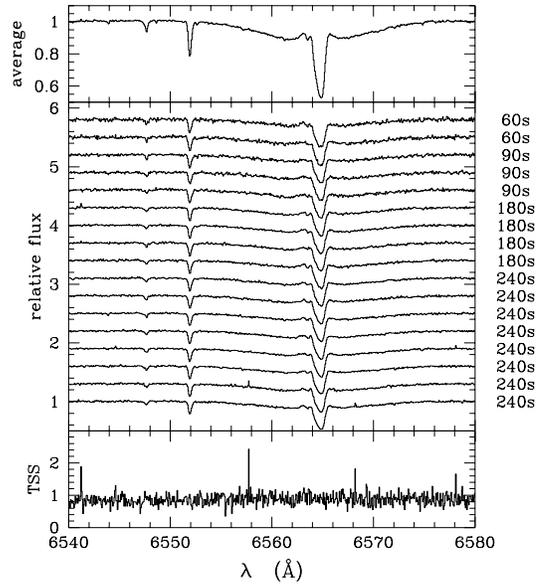,width=\columnwidth}}}
\caption{The 17 spectra that were obtained around \ha during the December run.
The spectra are continuum
normalized, each spectrum is plotted with an offset of 0.3
for comparison. The Temporal
Sigma Spectrum (see text for details) is also plotted. Note that the small
spikes in the individual spectra are immediately visible in the $TSS$}
\label{art8fig1}
\end{figure*}

In Fig.~\ref{art8fig1} all individual spectra around \ha obtained
December 1993 are
plotted.
In order to decide whether there are any variations in the spectra
we adopt the simple, yet powerful
statistical formalism presented by Fullerton (1990) and Henrichs~\etal
(1994).  With this method the variability can be expressed in a
temporal variance spectrum ($TVS$, Eq.~\ref{art8eq1}):

\begin{equation}
(TVS)_{\lambda} \approx \frac{1}{N-1} \sum_{i=1}^{N}\left(
\frac{ F_{i}(\lambda) -
F_{\rm av}(\lambda)}{\sigma_{i}(\lambda)}\right)^{2}
\label{art8eq1}
\end{equation}

where $N$ is the number of spectra, $F_{\rm av}(\lambda)$ represents the
constructed average spectrum, $F_{i}(\lambda)$ the individual spectra, and
$\sigma_{i}(\lambda) =  F_{i}(\lambda) /(SNR)$ of each individual pixel of
the spectra.

Following Henrichs~\etal (1994) and Fullerton (1990) the temporal sigma
spectrum ( $TSS = \sqrt{TVS}$) is calculated.  This quantity represents
approximately ($\sigma_{obs}/\sigma_{\rm av}$), that is the standard
deviation of the variations of the individual spectra with respect to
the average spectrum divided by the standard deviation of the average
spectrum.  If no significant variations are present in the spectra, the
value will be close to one, significant deviations are directly
represented in units of the noise level, that is to say a peak ``Temporal
Sigma Spectrum'' of three corresponds to a variability at a 3$\sigma$
level.

\begin{figure}
\centerline{\hbox{\psfig{figure=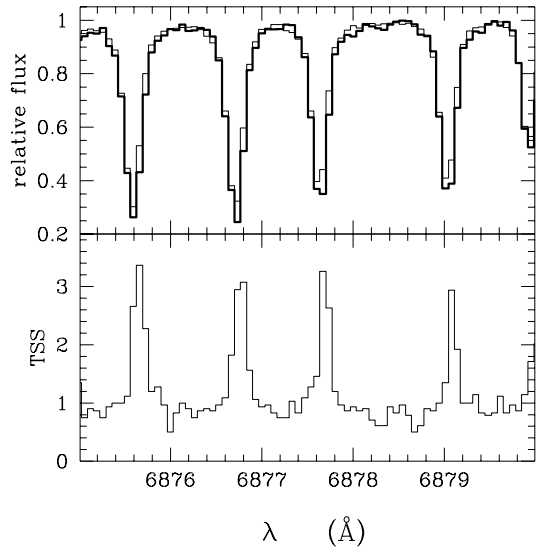,width=\columnwidth}}}
\caption{An overplot of the first spectrum (thick line, airmass 1.47) and the
last
spectrum (thin line, airmass 1.16) taken
around 6875~\AA~ in December is shown in the upper panel. The lower panel
shows the Temporal Sigma Spectrum. It can be seen that the
variable telluric absorption lines show up as significant peaks in
the $TSS$}
\label{art8fig2}
\end{figure}

The individual spectra and
the resulting $TSS$ are plotted around \ha in Fig.~\ref{art8fig1}.
It is clear from the temporal sigma spectrum that no significant
variability is present.
The spikes in the $TSS$ spectrum correspond
to cosmic ray hits that resulted in higher count levels and
consequently higher signal-to-noise ratios. These cosmic ray events
are still a bit present in the spectra, and provide an independent check
whether the method is indeed capable of detecting variability.
Another way to demonstrate the adequacy of the method can be found in
Fig.~\ref{art8fig2}, where we have plotted the first and last taken spectrum
around
6875~\AA~ during the December run.  The effect of the decreasing
airmass on the strength of telluric absorption lines is seen.

\begin{figure}
\centerline{\hbox{\psfig{figure=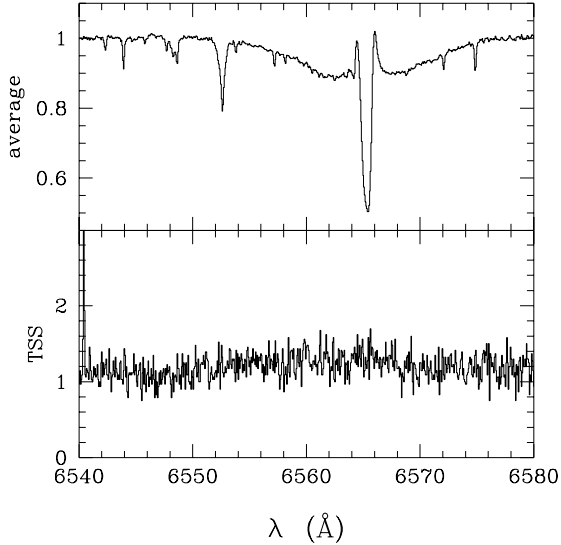,width=\columnwidth}}}
\caption{The summed spectrum and the resultant Temporal Sigma Spectrum of the
observations  obtained in February 1994.
The slight curvature of the $TSS$  arises from the fact that the
normalization of the spectrum was difficult, leading to small variations
of the continuum in the wings of the \ha line}
\label{art8fig3}
\end{figure}

The same procedure was carried out for the February observations.
The average spectrum and the resulting $TSS$
are shown in Fig.~\ref{art8fig3}.
No
variability can be found either in the February run.  The shallow shape
of the $TSS$ arises from the fact that the normalization of the spectrum
was difficult, leading to a small variation of the continuum in the
wings of the \ha line.

Similar calculations were performed for spectra with the same
integration times, and for the rest of the UES spectrum.  Except for
strong telluric absorption lines (see Fig.~\ref{art8fig2}) no variations
are present.

\subsection{Time scales of months}
\label{art8secmon}

Average spectra were computed by adding all spectra before the continuum
correction, resulting in spectra with a high $SNR (\geq 300$ and $ \leq  175$
for the December and February observations respectively).  The radial
velocity is not significantly variable, the heliocentric velocity we
derive from strong metallic lines are 87\mpm2~\kms, and 84\mpm2~\kms\,
for the December and February spectra respectively.

\begin{figure}
\centerline{\hbox{\psfig{figure=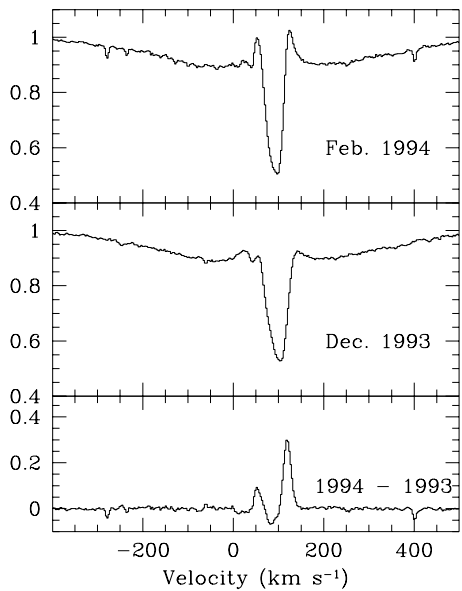,width=\columnwidth}}}
\caption{\ha profiles of HD~56126 at the two epochs. The difference spectrum is
shown in the lower panel}
\label{art8fig4}
\end{figure}

In Fig.~\ref{art8fig4} the \ha lines at both occasions are shown.
The profiles are
different; whereas in the December spectrum the \ha profile
shows a central absorption at 103~\kms, it has shifted to 94~\kms in
February, the peaks around the \ha absorption have increased in
strength.  The true shape of the emission is hard to determine, but
considering the red-shifted center of the \ha absorption, there
appears to be a rather broad emission component blue-shifted by more than
20~\kms.

Not only the \ha lines differ, a preliminary comparison of the
individual spectra shows that all spectral lines are broader by more
than 40\% at their full width half maximum, while the depression
relative to the continuum decreased in the February spectra relative to
the December spectrum.  Interestingly, except for the strongest metallic
lines (depression $\leq 50 \%$) where a blue-shifted absorption wing is
visible, all lines appear symmetric.  The interstellar component in the
profiles of the NaI D1 \& D2 lines do not change in velocity nor in
shape between the two spectra, indicating that changes in other lines
are not due to observational errors.  For the FeI and FeII lines we note
that the equivalent width of the FeI lines decreased,
while the equivalent width of the FeII lines has increased.  The fact
that the ionization degree of Fe has changed indicates an increase of
the effective temperature, the broader lines are probably the result of
a higher surface gravity.  The deeper central absorption in the \ha
line is also due to an increase of the effective temperature which gives
a higher population of the $n=2$ level for hydrogen according to the
Boltzmann law.

\section{Discussion}

In  Sect.~\ref{art8secmin} it was shown that no short term
variability is found in the \ha line of HD~56126.

Before we discuss any explanation for this observational result, let us
first review the data presented by TT93.  They presented five
spectra of HD~56126
taken with the 74 inch Okayama telescope.  The
spectra were obtained within 47 minutes, with 6 minutes the
shortest time interval between two observations.
The wavelength resolution is about 0.13~\AA~ (6~\kms at \ha),
comparable to the spectra presented here.
{}From their plots we find for their best spectrum a peak-to-peak noise of
approximately 17\% of the continuum level and of about  30\% for the
other spectra.  We estimate their $SNR$ at about 10-20.
Of the five spectra three show an absorption profile,  only two show the
variation noticed by TT93.  One of these shows a red emission peak
while another spectrum shows a blue
peak, which they claim to be ``drastic changes of the \ha profile in
the order of ten minutes''.
In fact, these are the only spectra  that show noticeable
variations.
The signal-to-noise of their spectra is at most 20, so
that a 1$\sigma$ variation has to be at least a 5\%  change for an
unresolved line. The (variable) emission components hardly exceed this
value.
Furthermore, the strong SrI line at 6550.2~\AA~ ($W  = 7$~m\AA,
depression to $0.8 \times$ continuum) is present in only two of their
spectra. This further suggests that the observed changes
may be artifacts of the noise rather than true changes in the
spectrum.

In Sect.~\ref{art8secmon}
 it was argued that, due to the broadening of most stellar
lines and the change in ionization degree of Fe, the star has become
hotter over a time interval of 65 days. This indicates that the star is
variable, consistent with the findings of Bogaert
(1994), who found photometric variability of HD~56126.

It is still hard to identify the mechanism responsible for the
emission observed in \ha.
It can be due to mass-loss, where the stronger emission in the February
spectrum is simply due to the increased temperature of the star, it
could also be due to a shock wave ploughing through the photosphere.
It may be worthwhile to monitor post-AGB stars on time scales of days and
weeks to appreciate the true temporal changes in their spectra.

Finally we consider whether the absence of any short term
variability can be expected based on simple arguments;
in principle the fastest time scale in which a wave can propagate in a
stellar atmosphere is the
scale height $h$ divided by the speed of sound $v_{s}$:

\begin{equation}
h  = \frac{k T R^{2}}{\mu m_{H}MG} ~{\rm [m]}
\end{equation}

\vspace{1 ex}

\begin{equation}
v_{s}  \sim \sqrt{\frac{k T}{\mu m_{H}}} ~{\rm  [km ~s^{-1}]}
\end{equation}

\noindent
With $M$ the mass of the star and $R$ its radius, $\mu$ is the mean
molecular weight which is  taken to be 1.
Adopting typical parameters for a F-type
post-AGB star, $M = 0.6$~\msol, and $R=60 R_{\odot}$, we obtain a
shortest possible time scale of 1.8 days.

One can also  estimate the pulsation period of a typical
post-AGB star:\\

\begin{equation}
\tau_{P} = Q \times \sqrt{
\left( \frac{R_{*}}{R_{\odot}} \right)^{3}
\left( \frac{M_{\odot}}{M_{*}} \right)} ~{\rm [days]}
\end{equation}

Where $Q$ is the pulsation constant in days. Adopting the same stellar
parameters as above, and noting that $Q$ ranges from 0.04  for
classical Cepheids to 0.16 for W-Virginis stars (Allen 1973)
we obtain an expected period of 30 -- 96 days.
This simple exercise tells us
that variations on time scales shorter than days are not really to be
expected in post-AGB stars.  It is consistent with the change observed
over a period of two months.

\section{Conclusions}

We observed the post-AGB star HD~56126 (F5I) at short time intervals in
order to look for very short term variations of the \ha line. Despite
the high quality of the data we were not able to find significant
variations.
The report by Tamura and Takeuti (1993) that the line is variable within
minutes could not be confirmed. We believe that the
variability they claim is due to the low signal-to-noise ratio of their
spectra and not due to real variability.
Strong variations are found over the two month  interval, where the
central absorption of the \ha line has shifted and both peaks around
the central absorption have increased considerably.
It is shown that the star has increased its surface temperature over two
months, indicating that the underlying star rather than the
circumstellar environment changes, adding more validity to pulsational
models that account for observed changes in spectra of post-AGB stars.
\newline

{\it Acknowledgements} We wish to express our gratitude to Ren\'e Rutten
for carrying out the service observations.  Ton Schoenmaker is greatly
acknowledged for introducing RDO to the IRAF reduction software package.
We enjoyed the enlightening discussions with   Peter van Hoof, Rens
Waters, Henny Lamers and Norman Trams.  Griet Van de Steene is
acknowledged for her comments on an earlier version of the manuscript.
Finally we wish to thank the referee, Christoffel Waelkens for his
constructive remarks.

RDO and EJB receive financial support
under grants no.  782-372-031 and 782-371-040 from the Netherlands
Foundation for Research in Astronomy (ASTRON) which receives its funds
from the Netherlands Organization for Scientific Research (NWO).

The
William Herschel Telescope is operated on the island of La Palma by the
Royal Greenwich Observatory in the Spanish Observatorio del Roque de los
Muchachos of the Instituto de Astrof\'\i sica de Canarias.

This research
has made use of the Simbad database, operated at
CDS, Strasbourg, France.

\end{document}